\documentclass[a4paper,12pt,twoside]{article}
\usepackage{latexsym}
\usepackage{amsfonts}
\else\oddsidemargin25mm\evensidemargin25mm\marginparwidth25mm\fi
\begin{document}
\def\a{\alpha}
\def\b{\beta}
\def\g{\gamma}
\def\d{\delta}
\def\e{\epsilon}
\def\ve{\varepsilon}
\def\t{\theta}
\def\l{\lambda}
\def\m{\mu}
\def\n{\nu}
\def\pg{\pi}
\def\r{\rho}
\def\s{\sigma}
\def\t{\tau}
\def\c{\chi}
\def\p{\psi}
\def\o{\omega}
\def\G{\Gamma}
\def\D{\Delta}
\def\T{\Theta}
\def\L{\Lambda}
\def\Pg{\Pi}
\def\S{\Sigma}
\def\O{\Omega}
\def\pb{\bar{\psi}}
\def\cb{\bar{\chi}}
\def\lb{\bar{\lambda}}
\def\i{\imath}
\def\eq#1{(\ref{#1})}
\newcommand{\be}{\begin{equation}}
\newcommand{\ee}{\end{equation}}
\newcommand{\ba}{\begin{eqnarray}}
\newcommand{\ea}{\end{eqnarray}}
\newcommand{\ban}{\begin{eqnarray*}}
\newcommand{\ean}{\end{eqnarray*}}
\newcommand{\nn}{\nonumber}
\newcommand{\nin}{\noindent}
\newcommand{\fgl}{\mathfrak{gl}}
\newcommand{\fu}{\mathfrak{u}}
\newcommand{\fsl}{\mathfrak{sl}}
\newcommand{\fsp}{\mathfrak{sp}}
\newcommand{\fusp}{\mathfrak{usp}}
\newcommand{\fsu}{\mathfrak{su}}
\newcommand{\fp}{\mathfrak{p}}
\newcommand{\fso}{\mathfrak{so}}
\newcommand{\fl}{\mathfrak{l}}
\newcommand{\fg}{\mathfrak{g}}
\newcommand{\fr}{\mathfrak{r}}
\newcommand{\fe}{\mathfrak{e}}
\newcommand{\rE}{\mathrm{E}}
\newcommand{\rSp}{\mathrm{Sp}}
\newcommand{\rSO}{\mathrm{SO}}
\newcommand{\rSL}{\mathrm{SL}}
\newcommand{\rSU}{\mathrm{SU}}
\newcommand{\rUSp}{\mathrm{USp}}
\newcommand{\rU}{\mathrm{U}}
\newcommand{\rF}{\mathrm{F}}
\newcommand{\R}{\mathbb{R}}
\newcommand{\C}{\mathbb{C}}
\newcommand{\Z}{\mathbb{Z}}
\newcommand{\Hb}{\mathbb{H}}

\begin{titlepage}
\begin{flushright}
\hskip 5.5cm \hskip 1.5cm
\vbox{\hbox{CERN-TH/2002-314}\hbox{UCLA/02/TEP/27}\hbox{November,
2002}} \end{flushright} \vskip 3cm
\begin{center}
{\LARGE {No--scale $N=4$ supergravity coupled to Yang--Mills: the scalar potential and super-- Higgs effect}}\\
\vskip 1.5cm
  {\bf R. D'Auria$^{1,2,\star}$, S. Ferrara$^{3,4,5,\ast}$,M.A.Lled\'o$^{1,2,\star\star}$ and S. Vaul\`a$^{1,2,\circ}$} \\
\vskip 0.5cm
\end{center}
{\small $^1$ Dipartimento di Fisica, Politecnico di
Torino, Corso Duca degli Abruzzi 24, I-10129 Torino}\\
{\small $^2$ Istituto Nazionale di Fisica Nucleare (INFN) -
Sezione di
Torino, Italy}\\
{\small $^3$ CERN, Theory Division, CH 1211 Geneva 23, Switzerland,}\\
{\small $^4$ INFN, Laboratori Nazionali di Frascati, Italy}\\
{\small $^5$ Department of Physics and Astronomy, University of
California, Los Angeles, California, USA} \vskip 0.5cm
\begin{center}
e-mail: $^{\star}$ dauria@polito.it, $^{\ast}$
sergio.ferrara@cern.ch,
$^{\star\star}$ lledo@athena.polito.it., $^{\circ}$ vaula@polito.it \vskip 6pt
\end{center}
\vskip 3cm
\begin{abstract}
 We derive the scalar potential of the effective theory of type
 $IIB$
orientifold with 3-form fluxes turned on in presence of non
abelian brane coordinates.\\ $N=4$ supergravity predicts a
positive semidefinite potential with vanishing cosmological
constant in the vacuum of commuting coordinates, with a classical
moduli space given by three radial moduli and three RR scalars
which complete three copies of the coset  $(U(1,1+n)/U(1)\otimes U(1+n))$,
together with $6n$ $D3$--branes coordinates, $n$ being the rank
of the gauge group $G$. Implications for the super Higgs
mechanism are also discussed.
\end{abstract}

\end{titlepage}

\section{Introduction}
Recently it was shown \cite{D'Auria:2002tc} that the bulk sector
of the type $IIB$ orientifolds with 3--form fluxes turned on
\cite{fp}, \cite{kst} corresponds, for a generic choice of fluxes
which give a vanishing vacuum energy, to a gauged $N=4$
supergravity with a linear action of $SL(2)\times GL(6)$ on the
12 vector potentials (belonging to the representation ({\bf
2,6})) of $N=4$ supergravity coupled to six $N=4$ vector
multiplets
\cite{Andrianopoli:2002aq}.\\
An important ingredient is that, in order to obtain the correct
theory, only a $GL(6)$ subgroup of the global symmetry group
$SO(6,6)$ acts linearly on the gauge potentials, while other
$SO(6,6)$ transformations mix electric with magnetic field
strengths
\cite{Tsokur:1994gr}.\\
Fluxes turned on correspond, in the supergravity language, to a
gauging of translational isometries of $SO(6,6)$ in the
decomposition
\be\fso(6,6)=\fgl(6,\R)+\mathbf{15^{'+}}+\mathbf{15^-}\ee The
only fields which are charged under this gauging are the 15 RR
scalars whose dual have a charged coupling to the 12 vectors \be
\nabla_{\m}B^{\L\S}=dB^{\L\S}+f^{\L\S\D\a}A_{\D\a}\ee Here
$\L,\,\S,\,\D$ are $GL(6)$ indices, the axion $B^{\L\S}$ is an
antisymmetric tensor and the "charges" $f^{\L\S\D\a}$ are the
3--form (RR and NS) fluxes
\cite{D'Auria:2002tc}, \cite{fp}, \cite{kst}.\\
The corresponding $N=4$ supergravity sector has a scalar
potential given by \be V=\frac{1}{12}|F^{-\,IJK}|^2\ee where \be
\label{self}F^{-\,IJK}=\frac{1}{2}\left(F^{IJK}-
i\,^*F^{IJK}\right)\ee and \be \label{ariself}
F^{IJK}=L^{\a}f_{\a}^{IJK}\ee $L^{\a}$ parametrizes the
$SU(1,1)/U(1)$ coset of the NS and RR dilaton sector as follows:
let the generic element of $SU(1,1)/U(1)$ be given by $s$,
\be\label{cososcal} s=\begin{pmatrix}{\phi_1&\overline{\phi}_2\cr
\phi_2&\overline{\phi}_1\cr}\end{pmatrix}\ \ \ \ \ \
(\phi_1\overline{\phi}_1-\phi_2\overline{\phi}_2=1)\ee introducing
the 2-vector \be L^{\a}\equiv\begin{pmatrix}{L^1\cr
L^2\cr}\end{pmatrix}
=\frac{1}{\sqrt{2}}\begin{pmatrix}{\phi_1+\phi_2\cr-i(\phi_1-\phi_2)\cr}\end{pmatrix}\ee
\be L_{\a}\equiv\epsilon_{\a\b}L^{\b}\ee  the identity
$\phi_1\overline{\phi}_1-\phi_2\overline{\phi}_2=1$ becomes
\be\label{procione}
L^{\a}\overline{L}^{\b}-\overline{L}^{\a}L^{\b}=i\e^{\a\b}\ee To
relate this to the standard complex dilaton, \be
S=ie^{\varphi}+C\ee where $\varphi$ is the string dilaton and $C$
is its RR partner, let us set
\be\frac{\phi_2}{\phi_1}=z=\frac{-i+S}{i+S}\ee then we have \be
e^{K}=\frac{1}{2i(\overline{S}-S)}=\frac{e^{-\varphi}}{4}\ee so
that
\ba&&\phi_2=\frac{1}{2}[i(e^{\varphi}-1)+C]e^{-\frac{\varphi}{2}}\\
&&\phi_1=\frac{1}{2}[i(e^{\varphi}+1)+C]e^{-\frac{\varphi}{2}}\ea
Using the above relations, one easily shows that the potential
given in reference \cite{D'Auria:2002tc} agrees with the
potential recently discussed in reference \cite{Frey:2002qc} since
\be|\phi_2|^2=\frac{1}{2}(cosh\,\varphi-1+\frac{1}{2}e^{-
\varphi}C^2)\ee The gravitino mass matrix is given by
\cite{D'Auria:2002tc},
\cite{Tsokur:1994gr}\be\label{grashift}S_{AB}=
-\frac{i}{48}\overline{F}^{-IJK}(\G_{IJK})_{AB}\ee This is a
no--scale model \cite{Cremmer:1983bf}, \cite{Ellis:1983sf} in
that the contribution of $|S_{AB}|^2$ to the gravitational
potential exactly cancels (even away from the extremum) with the
four goldstinos which come from the ${\bf \overline 4}$ in the
${\bf 6}\times {\bf {4}}={\bf 20}+{\bf \overline 4}$  $SU(4)$
decomposition of the gauginos of the six matter multiplets.

\section{Coupling to Yang--Mills matter}

The above discussion gives the result derived in reference
\cite{D'Auria:2002tc} which agrees with the potential obtained by
the bulk part of the action of type $IIB$ on $T^6/\mathbb{Z}_2$
orientifold \cite{fp}, \cite{kst}, \cite{Frey:2002qc}.\\ A less
obvious result is the coupling of this system to additional
Yang--Mills matter, that is in presence of additional $n$ vector
multiplets which are Lie algebra valued on some compact group
$G$. This result, in the superstring framework, comes from the
coupling of the Born--Infeld non abelian action to the gravity
sector of the ambient space. Some of these couplings have been
computed in the literature \cite{Myers:1999ps},
\cite{Polchinski:2000uf}, but their completion, to give an
effective theory with $N=4$ local supersymmetry, is not a
straightforward exercise.\\ This completion was obtained
\cite{Ferrara:2002bt} in the particular case of a $N=1$ sector
coupled to Yang--Mills, as one would obtain if all degrees of
freedom (in particular three massive gravitinos) in the partial
breaking $N=4\longrightarrow N=1$ were integrated out. The
result, with only one residual flux breaking $N=1\longrightarrow
N=0$, is a no--scale model with a particularly simple
structure.\\ The amazing fact of this result is that the non
abelian effective theory does not contain, for example, terms of
order $a^3$, where $a$ are the $D3$--brane coordinates, but only
a pure $a^4$ term, as in the pure Born--Infeld with non
gravitational back--reaction corrections. The cubic term is
predicted by an explicit calculation \cite{Myers:1999ps} but it
is known to vanish if the equation $F^{IJK-}=0$ is used. This
equation, among other things, stabilizes the (complex) dilaton in
terms of flux entries \cite{fp}, \cite{kst}.\\ To have a better
insight of these results we now present the full $N=4$ potential
where both fluxes and non abelian brane coordinates are present.
This potential is completely predicted by $N=4$ supergravity with
gauge group $\mathcal{G}$ of dimension $12+n$ ($n=dim\,G$) which
is the direct product of 12 translations with the compact Lie
group $G$: $\mathcal{G}=T_{12}\otimes G$. Note that this requires
a non standard symplectic embedding \cite{Gaillard:1981rj} of the
full $SL(2,\mathbb{R})\times SO(6,6+n)$ symmetry in $Sp(24+2n)$
group because $SL(2,\mathbb{R})$ acts linearly on the first 12
vectors, but acts as an electric--magnetic duality on the
remaining 2n field strengths and their dual. This embedding was
discussed in reference \cite{Tsokur:1994gr} and the subgroup
which acts on the gauge potentials is $GL(6)\times SO(n)$. In
this framework the scalar fields $B^{\L\S}$, $a^{\L}_i$
$(i=1\dots n)$ are treated as tensors of $GL(6)$ and they
complete the $GL(6)/SO(6)$ manifold to the full
$SO(6,6+n)/SO(6)\times SO(6+n)$. This formulation, in absence of
gauge coupling and fluxes, is related by a duality transformation
to the $N=4$ action constructed a long time ago in reference
\cite{Bergshoeff:1985ms}, \cite{deRoo:1985jh}. However, when the
fluxes and the non abelian couplings of $G$ are turned on, this
action is no longer equivalent to any of the previously proposed
actions and it allows to obtain a no--scale extended $N=4$
supergravity with non abelian gauge interactions.\\ The potential
can be computed from the fermion shifts modified by the charge
couplings. These shifts can be computed from the $N=4$ gauged
theory using superspace Bianchi identities 
\begin{eqnarray}
\d\p_{A\m}^{(shift)}&=&S_{AB}\g_{\m}\ve^B=-\frac{i}{48}
(\overline{F}^{IJK-}+\overline{C}^{IJK-})(\G_{IJK})_{AB}\g_{\m}\e^B\\
\d\c^{A\,(shift)}&=&N^{AB}\e_B=-\frac{1}{48}(\overline{F}^{IJK+}+\overline{C}^{IJK+})(\G_{IJK})^{AB}\e_B\\
\d\l^{I\,(shift)}_{A}&=&Z^{I\,B}_A\e_B=
\frac{1}{8}(F^{IJK}+C^{IJK})(\G_{JK})_{A}^{\,\,B}\e_B\\
\d\l^{(shift)}_{iA}&=&W^{\,\,\,B}_{iA}\e_B=\frac{1}{8}L_2\,q^{Jj}q^{Kk}\,c_{ijk}(\G_{JK})_{A}^{\,\,B}\e_B
\end{eqnarray}

Here $c_{ijk}$ are the structure constants of $G$, $C^{IJK}$ are
the boosted structure constants defined as \be
C^{IJK}=L_2q^{Ii}q^{Jj}q^{Kk}\,c_{ijk}\ee and
$q_{Ii}=E_{I\L}a^{\L}_i$ where $E_{I\L}$ are the coset
representatives of $GL(6)/SO(6)$.\\
The knowledge of the fermion shifts allows us to use the Ward
identity of supersymmetry in the Lagrangian \cite{lofaremo} to
compute the scalar potential which turns out to be \be
V=\frac{1}{12}|F^{IJK-}+C^{IJK-}|^2+\frac{1}{32}|L_2c_{ijk}q^{jJ}q^{kK}|^2\ee
Some comments are in order. If we set $q^{iI}=0$ (or commuting)
we retrieve the previous potential given in reference
\cite{D'Auria:2002tc}. On the other hand, if we set $F^{IJK-}=0$
we retrieve the standard potential (as for instance it comes from
the heterotic string compactified on $T_6$ \cite{Witten:1985xb},
\cite{Ferrara:1986qn}, \cite{Schwarz:mg}).\\
Interestingly, the first term contains an interference term \be
F^{IJK-}C_{IJK-}^*+cc\ee which is what was obtained from the
Born--Infeld action \cite{Myers:1999ps}. However, as it can be
seen, this term completes to a perfect square because of the
extra (pure gravitational) higher order $a^6$ term
$\frac{1}{12}|C^{IJK-}|^2$. This actually explains why,
integrating out the dilaton, such terms may disappear and this is
consistent with the $N=1$ reduction studied before
\cite{Ferrara:2002bt}.\\ Also note that the $a^6$ terms would not
always disappear if more supersymmetry remains unbroken, for
instance, integrating out the dilaton in the $N=4\longrightarrow
N=3$ reduction \cite{D'Auria:2002tc}, the non vanishing flux
$f_{ijk}=\e_{ijk}f$ just predicts that the pure holomorphic part
of $C^{IJK-}$, i.e. the singlet in the $SU(4)\longrightarrow
SU(3)$ decomposition ${\bf 10}\longrightarrow{\bf
6}+{\bf\overline{3}}+{\bf 1}$ should not be present. This is
indeed true as in the $N=3$ supergravity the $a^6$ terms are of
the form $|zz\overline{z}|^2$ \cite{Castellani:1985ka}, where
$z^A$ are holomorphic triplet coordinates in the splitting ${\bf
6}\longrightarrow{\bf 3}+{\bf\overline{3}}$ Note that a puzzle
seems to emerge on the fact that the Yang--Mills contribution to
the gravitino mass is proportional to \be S_{AB}\propto
c_{ijk}z^{Ci}z^{Dj}\overline{z}^k_{(B}\e_{A)CD}\ee where
$A,\,B,\,C$ are $SU(3)$ indices; this term is not holomorphic in
the $z^{Ai}$ coordinates, while it is holomorphic in the $N=1$
case \cite{Ferrara:2002bt}.\\ The resolution is the fact that the
supergravity transformations of the three $z^A$ (for a fixed
value of the $i$ index) and the four left--handed $\l_A,\,\l$
gauginos are \ba &&\d
z^A=\e^{ABC}\overline{\l}_B\e_C+\overline{\l}\e^A\\
&&\d\l_A=-i\e_{ABC}\g^{\m}\partial_{\m}z^B\e^C\\
&&\d\l=i\g^{\m}\partial_{\m}\overline{z}_A\e^A\ea So, if we pick
up $\e^A=(\e^1,0,0)$ as in $N=1$ reduction, we see that the
left--handed $N=1$ multiplets are \be
(z^3,\,\l_2)\quad(z^2,\,\l_3)\quad(\overline{z}_1,\,\l) \ee so
that we can identify $\overline{z}$ with the third holomorphic
coordinate. The remaining $\l_1$ is the $N=1$ gaugino.\\
The condition for the potential to have minimum with vanishing
cosmological constant is \ba&&\d\l_{iA}^{(shift)}=0\\
&&\d\c^{A(shift)}=0\\ &&\d\l_{IA}^{(20)(shift)}=0\ea The first
equation implies $c_{ijk}q^{iI}q^{jJ}=0$, that is the $q^{iI}$
are in the Cartan subalgebra of $G$. This also implies
$C^{IJK}=0$ and then the other equations imply $F^{IJK-}=0$.\\
The latter equation implies, for arbitrary values of the
gravitino masses, that the complex scalar in $L_2$ (coordinates
of $SU(1,1)/U(1)$) and eighteen of the twenty--one coordinates of
$GL(6)/SO(6)$ are stabilized. Twelve axions in $B^{\L\S}$ are
eaten by twelve vectors with a Higgs mechanism
\cite{D'Auria:2002tc}, leaving $6+6n$ ($n=Rank(G)$) flat
directions.\\
By integrating out the massive modes, the space is (at least
classically) the product of three $CP^{n+1}=U(1,1+
n)/U(1)\times
U(1+n)$ $\s$--model, as was shown in reference
\cite{Ferrara:2002bt}.

\section{N=2 examples}

A particular interesting case to study is also the reduction to
$N=2$, where both vector multiplets and hypermultiplets are
present and the general for of the non linear $\s$--model is
predicted to be \cite{Ferrara:2002bt}
\be\frac{U(1,1+n)}{U(1)\times
U(1+n)}\times\frac{U(2,2+n)}{U(2)\times U(2+n)}\ee This theory has
two flux parameters corresponding to a gauging of a two
translational isometry of the $\frac{U(2,2)}{U(2)\times U(2)}$
coset. They are gauged by the two vectors of the $n=0$
gravitational sector \cite{Andrianopoli:2002aq}. Special geometry
predicts \cite{Ferrara:1995gu} that the $N=2\longrightarrow N=1$
partial breaking is possible only if the
$\frac{U(1,1+n)}{U(1)\times U(1+n)}$ special manifold has a
symplectic section which cannot be derived from a prepotential
function $F(X)$. We now show that this is indeed the case. The
argument is very similar to the analogous example, studied in
reference \cite{Ceresole:1995jg} for the special cosets
$\frac{SU(1,1)}{U(1)}\times\frac{SO(2,n)}{SO(2)\times SO(n)}$.\\
For the $CP_{n+1}$ special cosets, the holomorphic prepotential
in the natural basis where the $U(n)$ symmetry is manifest is \be
F(X)=i(X^0X^1+X^aX^a)\quad a=2,\dots n+1\ee with symplectic
(holomorphic) sections  $(X^{\L},\,F_{\L})$, $\L=0,1,\dots n+1$ with
\be F_0=iX^1;\quad F_1=iX^0;\quad F_a=2iX^a\ee The corresponding
K\"{a}ler potential is \be K=
-log\,i(\overline{X}^{\L}F_{\L}-X^{\L}\overline{F}_{\L})=-log[-2(S+\overline{S}+2x^a\overline{x}^a)]\ee
where $S=\frac{X^1}{X^0}$ and $x^a=\frac{X^a}{X^0}$. Let us now
perform the following symplectic change of holomorphic section,
using a symplectic matrix \be\mathcal{S}=\begin{pmatrix}{A&B\cr
C&D\cr}\end{pmatrix}\ee where \be A=D=\begin{pmatrix}{{\bf
p}_{2\times 2}& {\bf 0}_{2\times n}\cr {\bf 0}_{n\times 2}&{\bf
1}_{n\times n} }\end{pmatrix};\quad -B=C=\begin{pmatrix}{{\bf
q}_{2\times 2}& {\bf 0}_{2\times n}\cr {\bf 0}_{n\times 2}&{\bf
0}_{n\times n} }\end{pmatrix}\ee\vskip 0.5cm \be{\bf p}_{2\times
2}=\begin{pmatrix}{1&0\cr 0&0}\end{pmatrix};\quad\quad{\bf
q}_{2\times 2}=\begin{pmatrix}{0&0\cr 0&1}\end{pmatrix}\ee

which satisfy the symplectic conditions

\be A^TD-C^TB=A^2+C^2=1;\quad A^TC=B^TD=0\ee We have in the new
basis
\be\begin{pmatrix}{\widetilde{X}\cr\widetilde{F}}\end{pmatrix}=
\begin{pmatrix}{A&B\cr
C&D\cr}\end{pmatrix}\begin{pmatrix}{X\cr F}\end{pmatrix}\ee where
\be
\widetilde{X}^{\L}=(X^0,\,-iX^0,\,X^a);\quad\widetilde{F}_{\L}=(iX^1,\,X^1,\,2iX^a)\ee
Since $\widetilde{X}^{\L}$ does not contain $X^1$ this basis is
singular and no prepotential $\widetilde{F}(\widetilde{X})$
exists. The matrix $\mathcal{N}_{\L\S}$ can be obtained from the
special geometry relations \cite{Ceresole:1995jg} \ba
&&F_{\L}=\mathcal{N}_{\L\S}X^{\S},\quad\quad\mathcal{D}_{\overline{l}}\overline{F}_{\L}
=\mathcal{N}_{\L\S}\mathcal{D}_{\overline{l}}\overline{X}^{\S}\nn\\
&&\mathcal{D}_{i}X^{\L}=\partial_iX^{\L}+\partial_iKX^{\L},
\quad\mathcal{D}_{i}F_{\L}=\partial_iF_{\L}+\partial_iKF_{\L}\ea
The vector kinetic matrix $\mathcal{N}_{\L\S}$ turns out to be
holomorphic and is given in this basis by
\be\widetilde{\mathcal{N}}_{\L\S}=(\widetilde{\mathcal{N}}_{ab},\,\widetilde{\mathcal{N}}_{a0},\,
\widetilde{\mathcal{N}}_{a1},\,\widetilde{\mathcal{N}}_{01},\,\widetilde{\mathcal{N}}_{00},
\,\widetilde{\mathcal{N}}_{11})\ee with
\ba&&\widetilde{\mathcal{N}}_{ab}=-2i\d_{ab},\,\widetilde{\mathcal{N}}_{a0}=2ix_a,\,
\widetilde{\mathcal{N}}_{a1}=-2x_a\nn\\ &&
\widetilde{\mathcal{N}}_{01}=x^ax_a,
\,\widetilde{\mathcal{N}}_{00}=iS-ix^ax_a,\,\widetilde{\mathcal{N}}_{11}=iS+ix^ax_a\ea
So that the vector kinetic term takes the form \be
Im(\mathcal{N}_{\L\S}F^{+\L}_{\m\n}F^{+\S\m\n})\ee The above
result show that the $N=2$ model is compatible with partial
breaking of $N=2$ supersymmetry to $N=1,\,0$ ($N=2$ models with
partial supersymmetry breaking have been considered in the
literature \cite{Cecotti:1984fn}, \cite{Zinovev:1986jv},
\cite{Zinovev:1986yg}, \cite{Fre:1996js}). The moduli space of
such $N=2\longrightarrow N=1\longrightarrow N=0$ theories is
given by three copies ofthe coset $\frac{SU(1,1)}{U(1)}$. One is the
original $\frac{SU(1,1)}{U(1)}$ in the vector multiplet sector,
while
$\left(\frac{SU(1,1)}{U(1)}\right)^2\subset\frac{U(2,2)}{U(2)\times
U(2)}$ comes from the quaternionic manifold. Note that this is an
extension of the minimal model based on the coset
$\frac{Usp(2,2)}{Usp(2)\times Usp(2)}$ of reference
\cite{Ferrara:1995gu}.\\ A detailed analysis of the above
situation will be given elsewhere.

\section*{Acknowledgements}
S.Ferrara.would like to acknowledge interesting discussions with A.Frey
and M.B. Schulz. R.D'Auria and S.Vaul\`a acknowledge F. Gargiulo and M.
Trigiante for useful observations.M.A. Lled\'o would like to thank the Department of Physics and Astronomy of UCLA
for its kind hospitality during the completion of this work.

  Work supported in part by the
European Community's Human Potential Program under contract
HPRN-CT-2000-00131 Quantum Space-Time, in which  R. D'Auria,M.A.Lled\'o and  S.
Vaul\`a are associated to Torino University. The work of S. Ferrara has
also  been supported by the D.O.E. grant DE-FG03-91ER40662, Task
C.

\end{document}